\documentstyle[12pt]{article}

\begin{document}
\title{Influence of $SU(2) \otimes U(1)$ singlet scalars on Higgs boson 
signal at LHC}
\author{N.V.Krasnikov \thanks{On leave of absence from INR, Moscow 
117312, Russia}  \\
CERN, Geneva 23, Switzerland}
\date{September,1997}
\maketitle
\begin{abstract}

We investigate a model of $SU(2) \otimes U(1)$ singlet scalars 
coupled to Standard Model. We point out that for the 
case of the maximal  mixing between singlet scalars and standard 
Higgs boson the Higgs boson production cross 
section is smaller than the standard Higgs boson production 
by factor $\frac{1}{N+1}$, where N is the number of 
singlet scalars. For the case of big mixing Higgs boson could be 
nonobservable at LHC. However, there is also 
a possibility for the observation of scalar singlets at LHC. 
As a quasirealistic example we discuss supersymmetric 
$SU(2) \otimes U(1)$ electroweak model with an additional singlet 
chiral superfield.
  
\end{abstract}
\newpage

One of the main tasks of existing and future collider experiments 
(LEP2, LHC, NLC,..) is 
the search for the Higgs boson. Its detection would be the 
confirmation of the spontaneous symmetry 
breaking mechanism and an experimental "proof" of the 
renormalizability of the electroweak interactions. 
For the two most popular models of electroweak physics, 
the Standard Model(SM) and its minimal supersymmetric 
generalization(MSSM) LEP2 and LHC will be able to discover 
the Higgs boson with a mass from 77 GeV(current standard 
Higgs boson mass limit \cite{1}) up to $10^{3}$ GeV(upper 
limit comes from requirement of  tree level 
unitarity \cite{2}). In the SM and MSSM the Higgs boson 
for $m_{h} \leq 400$ GeV 
(this bound comes from an analysis of 
high precision LEP1 data \cite{3}) is rather narrow 
that leads to sharply peaked resonance. The signal to 
background ratio in the resonance region is mostly determined by the 
experimental energy resolution. 
    
In this paper we investigate the influence of 
$SU(2) \otimes U(1)$ singlet scalar fields on the detection of the 
Higgs boson at LHC. Namely, we investigate the influence  
of the mixing between $SU(2) \otimes U(1)$ scalar singlets and 
the Higgs boson. We find that in an extreme case of maximal mixing 
between $SU(2) \otimes U(1)$ scalars and 
the Higgs boson the Higgs boson production cross section 
drops by factor $\frac{1}{N+1}$ (N is the number of scalar singlets) 
that could lead to nonobservability of the Higgs boson at 
LEP or LHC. For instance, for LHC for $N \geq 2$ it is 
possible to make Higgs signal nonobservable for an integrated 
luminosity $L = 10^{5} pb^{-1}$. However, there is also 
possibility for the observation of the scalar singlets at LHC. 
As a quasirealistic example we discuss a 
supersymmetric electroweak model with singlet chiral superfield. 
In the limit when the number of singlet 
scalar fields $ N \rightarrow \infty $ we find that the 
Higgs boson mass is continuously distributed. A model 
of the Higgs boson with continuously distributed mass has 
been proposed in ref. \cite{4}. It should be noted that the 
models with additional light $SU(2) \otimes U(1) $ 
singlet scalar boson have been discussed in the literature \cite{5}. 
In such models Higgs boson invisible decay modes into light 
singlet scalars result in  a modification of branching 
ratios that leads to slightly different strategy for the 
Higgs boson search. In recent paper of T.Binoth and J.J. van der 
Bij \cite{6} a model of light scalar singlets strongly coupled to the 
standard Higgs boson has been discussed. In such model Higgs 
boson decays mainly into light singlet scalar bosons and 
its decay width is not small that considerably affects the signal 
to background ratio $S/B$ making Higgs signal diluted. We consider 
slightly different situation when big mixing of the 
Higgs boson with scalar singlets results in the appearance of 
many "Higgs like" scalar bosons in the spectrum and the 
production cross section of "Higgs like" bosons is smaller 
than the standard Higgs boson cross section production 
that leads to the Higgs signal dilution. In considered model 
the Higgs bosons branching ratios are not modified 
compared to the standard Higgs boson case. 

Let us start with the simplest modification of the SM model, namely, 
we add to the SM an additional real scalar 
singlet $\phi (x)$. Take the effective potential in the form      

\begin{equation}
V = \lambda (H^{+}H - \frac{v^2}{2})^2 + k(H^{+}H -
\frac{v^2}{2})\phi + \frac{M^2\phi^2}{2} .
\end{equation}
After the spontaneous symmetry breaking $<H> = \frac{v}{\sqrt{2}}$ , 
$<\phi> = 0$ we find that in the 
unitare gauge $H(x) = (0, \frac{v + h(x)}{\sqrt{2}})$ the 
mass matrix for the Higgs field h(x) and singlet 
fields $\phi(x)$ takes the form

\begin{equation}
\hat{M} = m^2_h h^2(x) + M^2\phi^2(x) + 2 \delta m^2 h(x)\phi (x) ,
\end{equation}
where $m^2_h = 2\lambda v^2$ and $\delta m^2 = \frac{k v}{2}$. 
The eigenstates of the mass matrix (2 ) are 
the fields
\begin{equation}
h_1(x) = h(x) \cos(\delta) -  \phi(x) \sin(\delta) ,
\end{equation}

\begin{equation}
\phi_1(x) = \phi(x) \cos(\delta)  + h(x) \sin (\delta )  
\end{equation}
with eigenvalues
\begin{equation}
M_{h_{1},\phi_{1}} = \frac{m^2_h + M^2}{2} \pm 
\sqrt{\frac{(m^2_h - M^2)^2}{4} + (\delta m^2)^2}
\end{equation} 
and with mixing angle
\begin{equation}
\tan(2\delta) = \frac{2\delta m^2}{M^2 -m^2_h} .
\end{equation}
Original Higgs field $h(x)$ is expressed in terms of the 
$h_1(x)$, $\phi_{1}(x)$ fields as 
\begin{equation}
h(x) = h_1(x) \cos(\delta) +\phi_{1}(x) \sin(\delta)
\end{equation}

The main production mechanisms of the Higgs boson at LEP and LHC 
are the Higgs bremsstrahlung of a real(LEP2) or 
virtual(LEP1) Z-boson and gluon fusion 
correspondingly. As a consequence of the formula (7) we 
find that the production cross sectons of the $h_1$ and 
$\phi_{1}$ bosons are the same  as for the standard 
Higgs boson production with the corresponding mass except multiplicative 
factors $\cos^2(\delta)$($h_1$-boson) and 
$\sin^2(\delta)$($\phi_1$-boson). The branching ratios of the 
$h_1$ and $\phi_1$ 
bosons into fermion-antifermion pairs, 2 gluons or 2 photons are 
the same as for the SM Higgs boson. As a result of 
the additional factors $\cos^2(\delta)$, $\sin^2(\delta)$ the 
signal significance $S = \frac{N_S}{\sqrt{N_B}}$ will be smaller 
than for the case of the SM Higgs boson provided the mass 
difference between the $h_1$ and $\phi_{1}$ bosons is bigger 
than the detector mass resolution. To be concrete consider 
the Higgs boson production at LHC
\footnote{LEP1 data \cite{7} lead to the bound $\sin^{2}(\delta) 
 \leq 0.1$ for $m_{\phi_{1}} \leq 30$ GeV}. For the case 
100 GeV $ \leq m_h \leq $ 130 GeV the most promising signature 
for the search for the Higgs boson is 
\begin{equation}
pp \rightarrow (h \rightarrow \gamma \gamma) + ...
\end{equation}
For low luminosity stage with an integral luminosity 
$L = 3 \cdot 10^{4}pb^{-1}$ CMS detector will be able to 
discover the standard Higgs boson with the signal 
significance \cite{8} $6 \leq S \leq 9$. The ATLAS 
detector \cite{9} will have the similar signal significance 
as the CMS. The mass resolution of the diphoton pair 
invariant mass at low luminosity stage is  assumed to 
be $\delta m_{\gamma \gamma} \approx 0.5$ GeV. For the case of 
maximal mixing $\delta = \frac{\pi}{4}$ and for  
$M_{h_1} - M_{\phi_{1}} \geq 1 $ GeV as a result of nonzero mixing we shall 
have two resonances in the diphoton spectrum. 
 For the case of maximal mixing the Higgs boson production 
cross section decreases by factor 2 
and the signal significance of each of two 
Higgs like resonances will be between 3 and 4.5. According 
to standard convention 
Higgs boson is assumed to be discovered provided 
the signal significance is bigger than 5. So at low luminosity stage 
in the case  of maximal mixing the Higgs boson
 escapes from being discovered. At high luminosity stage for an 
integral luminosity $L = 10^{5}pb^{-1}$ the signal 
significance for CMS detector 
for  110 GeV $ \leq m_h \leq $ 140 GeV 
is estimated to  be $ 11 \leq S \leq 13$. At high 
luminosity stage two Higgs like resonances will be detected at $S \geq 5$ 
significance level. Therefore there is nonzero 
probability instead of the single Higgs boson to discover 2 Higgs bosons by the 
measurement of the diphoton spectrum. 

For the Higgs boson mass interval 140 GeV $ \leq m_h \leq $ 600 GeV 
the most promising signature for the Higgs 
boson detection is through the decay 
 \begin{equation}
h \rightarrow ZZ^{*}(Z) \rightarrow 4 leptons .
\end{equation}
For the Higgs boson mass 200 GeV $\leq m_h \leq $ 400 GeV 
the signal significance for an integrated luminosity 
$L = 10^{5}pb^{-1}$ is estimated to be $ S = 16 - 18$ \cite{8}. 
For such mass interval the maximal mixing will be detected 
at $S \geq 5 $ significance level. Moreover for such mass 
interval it is possible to detect two Higgs like resonances at 
$5\sigma$ level provided the mixing angle $\sin^2(\delta) \geq 0.32$. 

At LEP1 and LEP2 the effects of nonzero mixing are the same 
as at LHC provided the mass difference between the scalars $h_1$ and 
$\phi_{1}$ is bigger than the detector energy resolution 
$\delta E \sim 4$ GeV. Again  instead of the single Higgs boson we shall 
have two Higgs like bosons with production cross sections 
for the maximal mixing two times smaller the standard one. 
Therefore the signal significance 
$S = \frac{N_S}{\sqrt{N_B}}$ is decreased by factor 
two compared to the standard case. An increase 
of an integral luminosity $L$ by factor 4 just compensates 
the decrease of the Higgs boson production cross section.

Consider now the case of N additional $SU(2) \otimes U(1)$ singlet 
scalar fields. Take the effective potential in the form  
\begin{equation}
V = \lambda (H^{+}H - \frac{v^2}{2})^2 + k_i (H^{+}H - 
\frac{v^2}{2})\phi_i +\frac{1}{2}m^2_{ij}\phi_{i} \phi_{j} ,
\end{equation}
where $i,j = 2,3,...N+1$. The equations for the determination of 
the nontrivial minimum of the effective potential (10) have solution 
$<H> = \frac{v}{\sqrt{2}}$, $<\phi_{i}>  = 0$. In the unitare gauge 
$H = (0, \frac{v + h(x)}{\sqrt{2}})$ the mass matrix for the Higgs field 
$\phi_{1}(x) \equiv h(x)$ and singlet fields 
$\phi_{i}(x)$ $(i = 2,3,..N+1)$ has the form 
\begin{equation}
\hat{M}^2 = m^2_{ij}\phi_{i}(x)\phi_{j}(x) ,,
\end{equation}
where $i,j =1,2...N+1$, $m^2_{11} = 2\lambda v^2$, 
$m^2_{1i} = m^2_{i1} = \frac{k_{i}v}{2}$. After the diagonalization 
the mass matrix (11) takes the form 
\begin{equation}
\hat{M}^{2} = M^2_i(\phi^{'}_{i})^2 ,
\end{equation}
where $\phi^{'}_{i} = O_{ij}\phi_{j}$, 
$\phi_{i} = O_{ji}\phi^{'}_{j}$ and $O_{ij}O_{ik} = \delta_{jk}$. 
Parameters $m^2_{ij}$ are arbitrary so 
in general orthogonal matrix $O_{ij}$ can have 
arbitrary form. Consider as an extreme example  the case of maximal mixing, i.e.
\begin{equation}
O_{1,1} = O_{2,1} = ... = O_{N+1,1} = \frac{1}{\sqrt{N+1}} . 
\end{equation}
For the mixing matrix (13) the original Higgs field 
$h(x) \equiv \phi_{1}(x)$ is a superposition of the of the fields 
$\phi^{'}_{i}(x)$ with 
definite mass
\begin{equation}
h(x) = \frac{1}{\sqrt{N+1}} \sum_{i=1}^{N+1}\phi^{'}_{i}
\end{equation}
Again as in the case of $N=1$ one can find that 
the branching ratios for the Higgs like bosons  
$\phi^{'}_{i}(x)$ are the same as for the 
standard Higgs boson but their  production cross section 
  drops by factor $\frac{1}{N+1}$ 
compared to the standard Higgs boson case. For the Higgs boson mass range 
 interval 100 GeV $ \leq m_h \leq $130 GeV and for an integrated 
luminosity $L = 10^{5}pb^{-1}$ the signal significance in an 
extreme case of maximal mixing is $2.7 \leq S \leq 4.1$ for $N=2$. 
So the Higgs boson will escape from being detected. 
However for total luminosity $L = 4 \cdot 10^{5} pb^{-1}$ 
the signal significance is bigger than $S \geq 5.4 $ and 3 Higgs like bosons 
could be discovered. Similar analysis works for the Higgs boson mass range 
interval 140 GeV $ \leq m_h \leq 400 $ GeV, where the most promising 
signature  is the use of the Higgs boson 
decay into 4 leptons (9).
As it follows from the LEP1 data \cite{7} it is possible for the Higgs like 
bosons with masses $ m \leq 30 $ GeV to escape from the detection 
for the maximal mixing case provided $N \geq 10$.
  
As a toy example consider  the following mass spectrum of the 
$\phi_{i}^{'}(x)$ fields:
\begin{equation}
M^2(i) = M^2_0 + \frac{\delta\cdot i}{N+1} ,
\end{equation}
where i = 1,2...N+1. In the limit $N \rightarrow \infty $ 
we shall have the Higgs boson with continuously 
distributed mass $M_0^2 \leq M^2_h \leq M^2_0 + \delta$. 
For $\delta \sim M^2$ Higgs boson has big decay 
width and escapes from the detection.  
Such Higgs boson with continuously 
distributed mass looks like broad resonance with decay width 
$\Gamma = \sqrt{M_0^2 + \delta} - M_0$ and 
a mass $M^2_0 +\frac{\delta}{2}$. The branching ratios for 
such "modified" Higgs boson coincide with the 
corresponding branching ratios of the standard Higgs 
boson unlike to the model of ref. \cite{6} 
where the big decay width of the Higgs boson is due to Higgs boson 
decays into invisible modes. 
Such models with 
continuously distributed Higgs boson mass have been discussed in ref. \cite{4}. 

As a quasirealistic example consider 
NMSSM(NMSSM = MSSM + chiral $SU(2) \otimes U(1)$ singlet 
superfield) \cite{10}. The superpotential of the NMSSM contains a term 
$\delta W = \lambda \sigma \bar{H} H$ in the superpotential 
that relates the 
$SU(2) \otimes U(1)$ singlet chiral superfield $\sigma$ 
with chiral superhiggs fields 
$H$ and $\bar{H}$. For small coupling constant 
$\lambda$ singlet superfield $\sigma$
effectively decouples from 
the MSSM superfields and most of the NMSSM 
predictions(spectrum, branching ratios, 
cross sections...)  coincide with the corresponding MSSM predictions. 
The general form of  the soft breaking terms 
for complex scalar singlet field 
$\sigma(x) = \frac{\sigma_1(x) + i\sigma_2(x)}{\sqrt{2}}$ 
is
\begin{equation}
V_{soft,\sigma} = \frac{m^2_1\sigma^2_1}{2} + 
\frac{m^2_2\sigma^2_2}{2} + m^2_{12}\sigma_{1} \sigma_{2} + 
(k_1\sigma_{1} + k_2\sigma_{2})\bar{H}H +c_1\sigma_{1} +c_2 \sigma_{2} 
+ ...
\end{equation}
The nondiagonal terms in (16) lead to  nonzero 
mixing between the singlet fields $\sigma_1(x)$, 
$\sigma_2(x)$ and the lightest Higgs field h(x). 
It is possible to choose parameters of the model 
such that the mixing is the maximal one, i.e. 
\begin{equation}
h(x) = \frac{1}{\sqrt{3}}(\phi_1(x) + \phi_2(x) + \phi_3(x)) ,
\end{equation} 
where the fields $\phi_i(x)$ have masses $M_i$. As it has been 
discussed before for 
such case we shall have Higgs boson signal 
dilution compared to the standard case.

To conclude, in this paper we have studied the influence of 
scalar singlets on the Higgs boson detection at LHC. We have 
found that big mixing of singlet fields with Higgs boson field 
can  lead to the nonobservation of the Higgs boson at LHC or to 
the observation of the additional Higgs like states.  
     
I am indebted to the collaborators of the INR theoretical department 
for discussions and critical comments. I thank CERN TH Depatment 
for the hospitality during my 
stay at CERN where this paper has been finished. 
The research described in this 
publication was made possible in part by Award No RP1-187 
of the U.S. Civilian Research and Development 
Foundation for the Independent States of the Former Soviet Union(CRDF).

Note added. After this paper has been finished I became aware of 
paper by Annindya Datta and Amitava Raychandhuty, hep-ph/9708444, 
where the effects of mixing between singlet and Higgs fields 
have been studied.


\newpage
\begin{thebibliography}{99}
\bibitem{1} A.Sopczak. Talk given at the international conference \\
Non-Accelerator searches for new physics, Dubna, 7-11 july 1997.
\bibitem{2} B.W.Lee, C.Quigg and C.B.Thacker, Phys.Rev. {\bf D10}(1974)1145.
\bibitem{3} J.Ellis, Phenomenology of LEP2 physics, CERN-TH/97-131.
\bibitem{4} N.V.Krasnikov, Phys.Lett.{\bf B325}(1994)430.
\bibitem{5} R.E.Schrock and M.Suzuki, Phys.Lett.{\bf B100}(1982)250;\\
L.F.Li, Y.Liu and L.Wolfenstein, Phys.Lett.{\bf B159}(1985)45;\\
D.Chang and W.Keung, Phys.Lett.{\bf B217}(1989)238;\\
N.V.Krasnikov, Phys.Lett.{\bf B291}(1992)89;\\
A.S. Joshipura and J.W.F.Valle, Nucl.Phys.{\bf B397}(1993)105.
\bibitem{6} T.Binoth and J.J. van der Bij, Z.Phys.{\bf C75}(1997)17.
\bibitem{7} O.Adriani et al.,, Phys.Lett.{\bf B294}(1992)457; \\
D.Buskulic et al., Phys.Lett.{\bf B313}(1993)312.
\bibitem{8} CMS, Technical Proposal, CERN/LHCC/94-38 LHCCP1, 15 december 1994.
\bibitem{9} ATLAS, Technical Proposal. CERN/LHCC/94-43 LHCCP2, 15 december 1994.
\bibitem{10} J.Ellis et al., Phys.Rev.{\bf D39}(1989)844;\\ 
L.Durand and J.L.Lopez, Phys.Lett.{\bf B217}(1989)463; \\
S.F.King and P.L.White, Phys.Rev.{\bf D52}(1995)4183; \\
M.Drees, Int.J. Mod.Phys.{\bf A4}(1989)3635;\\
U.Ellwanger, M.Rausch de Traubenberg and C.A.Savoy, Phys.Lett{\bf B315}(1993)331;\\
Z.Phys.{\bf C67}(1995)665; Nucl.Phys.{\bf B492}(1997)21.


    
\end{thebibliography}
\end{document}